# Room temperature single photon emission from oxidized tungsten disulphide multilayers

*Toan Trong Tran,[1,*] Sumin Choi,[1] John A. Scott,[1] Zai-quan Xu,[1] Changxi Zheng,[2,5] Gediminas Seniutinas,[1] Avi Bendavid,[3] Michael S. Fuhrer,[4,5] Milos Toth[1,^] and Igor Aharonovich[1]*

T. T. Tran, S. Choi, John A. Scott, Dr. Zaiquan Xu, Dr. G. Seniutinas, Prof. M. Toth, Prof. I. Aharonovich
School of Mathematical and Physical Sciences, University of Technology Sydney, Ultimo, NSW, 2007, Australia

E-mail: trongtoan.tran@student.uts.edu.au

Dr. C. Zheng
Department of Civil Engineering, Monash University, Clayton, VIC, 3800, Australia

Dr. A. Bendavid
Plasma Processing & Deposition Team, CSIRO Manufacturing Flagship, Australia

Prof. M. S. Fuhrer
School of Physics and Astronomy, Monash University, Clayton, VIC, 3800, Australia

Dr. C. Zheng, Prof. M. S. Fuhrer
Monash Centre for Atomically Thin Materials, Monash University, Clayton, VIC, 3800, Australia

**Abstract**

Two dimensional systems offer a unique platform to study light matter interaction at the nanoscale. In this work we report on robust quantum emitters fabricated by thermal oxidation of tungsten disulphide multilayers. The emitters show robust, optically stable, linearly polarized luminescence at room temperature, can be modeled using a three level system, and exhibit moderate bunching. Overall, our results provide important insights into understanding of defect formation and quantum emitter activation in 2D materials.

Two-dimensional (2D) materials have recently emerged as a promising platform for optoelectronics[1] and nanophotonics[2] owing to their atomically thin lattice structure and weak van der Waals interlayer interactions. Among these layered materials, transition metal dichacogenides (TMDs) such as molybdenum disulfide ($MoS_2$), tungsten disulfide ($WS_2$), molybdenum diselenide ($MoSe_2$), and tungsten diselenide ($WSe_2$) have garnered the most research interest thanks to their direct bandgaps (for monolayers), large carrier mobility, gate-tunability and mechanical flexibility.[3] The recent discoveries of single photon sources in the form of quantum dots in $WSe_2$[4] and $MoSe_2$[5] have further spurred intensive research into the use of TMDs for quantum information and processing. The 2D nature of these crystals and the know-how gained from work on manipulation of graphene, enabled rapid engineering of electrically triggered quantum light emitting diodes[6]

However, due to the shallow binding energies of several millielectron volts (meV), these quantum dots exhibit single photon emission only at cryogenic temperature. Furthermore, the nature of the quantum emitters is still under debate and their spectral filtering is challenging due to close proximity to excitonic lines of the host materials. On the other hand, room temperature (RT) single photon emission was recently observed from monolayers, few layers and bulk hexagonal boron nitride (hBN)[7, 8]. hBN has a wide bandgap of almost 6 eV and can therefore host a variety of localized defects with deep states that enable single photon emission at room temperature. In this letter, we report on quantum emission observed from annealed $WS_2$ multilayers. We show that annealing at a temperature of 550°C gives rise to partial oxidation of the flakes and to formation of localized stable optically active defects. We characterize the nature of the emissions and propose several models for the origin of the emitters.

Commercial solvent-exfoliated tungsten disulfide samples (Graphene Laboratories) were dropcast on a 1 $cm^2$ native oxide (a few nm thick $SiO_2$) silicon chip. The chip was annealed in a tube furnace (Lindberg/Blue M) at 550°C for 30 min under 1 Torr of flowing Argon and subsequently cooled down to room temperature before being subject to photoluminescence measurement. Similar to other TMDs, crystal structures of tungsten disulfide can be studied using Raman spectroscopy with three typical vibrational phonon modes $E_{2g}$, $A_{1g}$ and 2LA. While the first two arise from in-plane stretching optical phonon modes at Brillouin zone center, the latter come from a longitudinal acoustic mode.[9] Figure 1a shows a typical optical image of the annealed $WS_2$ multilayers on a native oxide silicon substrate (no difference was seen before and after the annealing process). We conducted Raman scattering measurement of the sample before and after thermal annealing at 550°C in an inert environment. A Raman spectrum from a high-quality monolayer $WS_2$ sample is used as a reference. Figure 1b presents Raman scattering plots for the three samples. Both the monolayer (red trace) and the pristine multilayers (green trace) show clear $E_{2g}$ and 2LA modes (grey highlighted), and $A_{1g}$ mode (orange highlighted), confirming the characteristic lattice vibrations of $WS_2$.[9] The annealed $WS_2$ sample, however, does not show these lattice vibration characteristics, implying that the flakes underwent compositional changes upon annealing. To elucidate the composition of the annealed $WS_2$ sample, we employed X-ray photoelectron spectroscopy (XPS). Figure 1c, 1d and 1e show photoelectron binding energies at $O_{1s}$, $S_{2p}$ and $W_{4f}$ regions, respectively. The three spectra suggest the presence of $WO_x$ and $WO_xS_y$ phases which form during the annealing process of $WS_2$, likely as a result of traces of residual water and oxygen molecules in the quartz tube. This behavior is consistent with previous studies on oxidation of $WS_2$.[10]

To characterize the annealed flakes further, we perform micro-photoluminescence (μPL) measurements using a typical home-built confocal microscope equipped with a high numerical aperture objective (NA = 0.9). The excitation of the samples is performed using a 532 nm continuous wave (CW) laser, and the collected signal is directed into two single photon detectors or imaged using a spectrometer. The details of the setup can be found elsewhere[7]. To examine the sample, we first conduct a confocal map scan over a 60 x 60 $\mu m^2$ area (Figure 2a). Several bright spots are observed in the confocal map, and their spectral properties are shown in Figure 2b. The PL spectra show single peaks at 600 nm (yellow trace), 650 nm (purple trace) and 730 nm (red trace), with peak widths broader than that of the excitonic emission line from the monolayer $WS_2$ sample (black trace). Note also that the yellow trace appears at higher energies than the excitonic transition of the $WS_2$ monolayer, further confirming that a new phase has formed. Figure 2c shows the three corresponding second order autocorrelation ($g^{(2)}(\tau)$) plots taken from the three centers described in Figure 2b. The $g^{(2)}(0)$ values are well below 0.5 at zero delay time, indicating that all three centers are indeed single photon sources.[13, 14] For convenience, we name the three emitters S1 (red), S2 (yellow) and S3 (purple). Note that emitters S1 and S3 exhibit similar spectral and temporal properties (i.e. no bunching at the same excitation power, and FWHM of 31 nm and

19 nm, respectively) while emitter S2 exhibits bunching and a much broader emission spectrum. To measure the excited state lifetimes of these color centers, we employed time-resolved PL with a 512 nm pulse laser (100 ps pulse width and 10 MHz repetition rate) as an excitation source. By using double exponential fitting, we obtained excited state lifetimes of 3.5, 4.6, and 4.4 ns for emitter S1, S2 and S3, respectively.

We proceed with further detailed characterization of emitter S1. A fluorescence saturation curve is recorded as a function of excitation power, and the data are shown in Figure 3a. To fit the data, we employ a standard three level system that has a ground, excited and a long lived metastable state. Consequently, the data are fit using equation 1, yielding a saturated intensity, $I_\infty$, of 350 kcps at a saturation power, $P_{sat}$, of 1.9 mW (Figure 3a). This brightness is comparable with room temperature emitters in diamond, ZnO and SiC.[11, 12]

$$I = I_\infty \frac{P}{P + P_{sat}} \quad (Eq. 1)$$

We then conducted an excitation and emission polarization study of emitter S1. By using a half-waveplate for excitation polarization measurement, and both a half-waveplate and a polarizing filter for emission polarization measurement, we obtained plots of excitation (red open circles) and emission (blue open squares) polarization for emitter S1 (Figure 3b). Fitting the two curves with a $\cos^2(\theta)$, indicates that the center is only partially polarized. The visibility of excitation and emission polarization of the defect were determined using equation 2 to be 0.32 and 0.79, respectively.

$$VIS = \frac{I_{max} - I_{min}}{I_{max} + I_{min}} \quad (Eq. 2)$$

The fact that the emission polarization has higher visibility than excitation polarization probably suggests that the absorption dipole moment of the emitter was not well-aligned with the polarization state of the excitation laser whereas the emission dipole moment of the emitter was.

To understand the photophysics of the emitter further, we measured $g^{(2)}(\tau)$ as a function of excitation power. This measurement is helpful to understand important photophysical parameters of quantum emitters such as their excited state ($\tau_1$) and metastable state ($\tau_2$) lifetimes which must be known in order to estimate the quantum yield of the centers. Such a measurement is shown in

Figure 4a. At high power (> 2mW), the emitter exhibits photo-bunching, indicating that its electronic structure includes at least one metastable state.[13] The curves are, therefore, fitted with a three-level model following equation 3, where $\tau_1$ and $\tau_2$ are excited and metastable state lifetime, respectively.

$$g^{(2)}(\tau) = 1 - a\, e^{-\tau/\tau_1} + b\, e^{-\tau/\tau_2} \quad (Eq.\,3)$$

By plotting $\tau_1$ and $\tau_2$ as a function of excitation power, we arrive at the graph shown in Figure 4b. The data are well fitted with single exponential,[14] yielding $\tau_1$ and $\tau_2$ values of 4.5 and 9.3 ns, respectively. It must be noted that the excited state lifetime obtained by this method is in relatively good agreement with that from the time-resolved PL measurement.

Finally, we discuss the optical stability of these emitters at room temperature. While the bandgap of a pristine monolayer $WS_2$ is around ~ 2 eV, the bandgap of $WO_xS_y$ can be substantially higher, up to 3.6 eV[15] in the case of $WO_x$. These higher bandgaps can facilitate localized defects with ground and excited state energies within the bandgap (see figure 4(c) and corresponding transitions that can be driven using a 2.4 eV green laser. The defect is likely a deep trap, as it is not thermally ionized and its quantum optical properties are preserved at room temperature. Similarly, zinc oxide (ZnO) that has a comparable bandgap of 3.2 eV also exhibits quantum emitters at room temperature[12, 16].

To summarize, we identified several quantum emitters in annealed $WS_2$ multilayers. Raman and XPS measurements suggest a phase transition to a $WO_x$ or $WO_xS_y$ phase via oxidation. The new layers host optically stable quantum emitters. The origin of the emitters is likely a deep trap defect state within the bandgap of the $WO_xS_y$. The quantum emitters show relatively high brightness and a short excited state lifetime suitable for photonic applications. A detailed study of various tungsten oxide materials is needed to understand the origin and the chemical structure of the defects. However, even at this point our results emphasize the breadth of emitters available within 2D materials and the need for further studies using high resolution optical techniques.

**Author Information**

*Corresponding authors: trongtoan.tran@student.uts.edu.au

*Notes: The authors declare no competing financial interest.

**Acknowledgements**

The work was supported in part by the Australian Research Council (Project Number DP140102721 and DP150103837), FEI Company and by the Air Force Office of Scientific Research, United States Air Force. I. A. and C. Z. are recipients of an Australian Research Council Discovery Early Career Research Awards (Project Number DE130100592 and DE140101555, respectively).

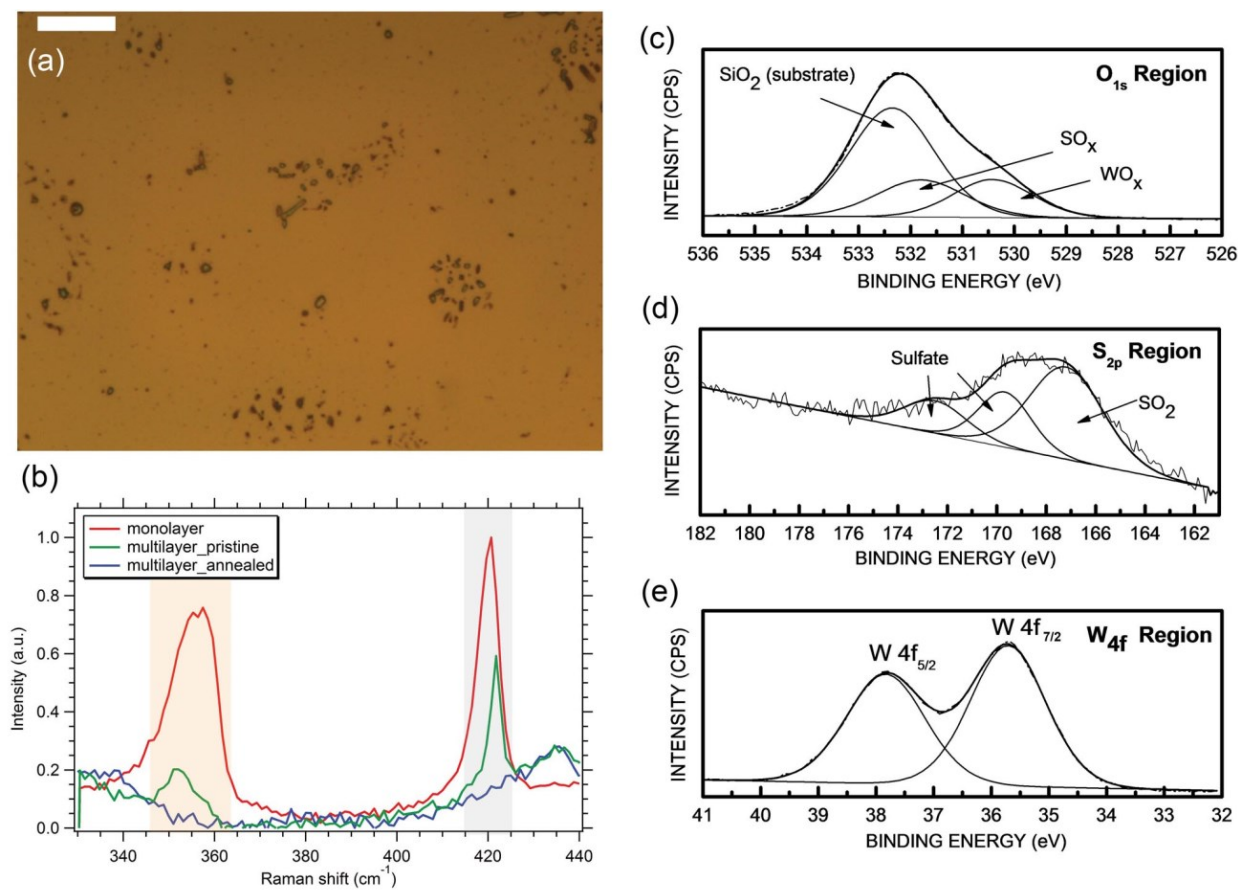

**Figure 1. Characterization of the multilayers.** (a) Optical image of annealed $WS_2$ multilayers. The scale bar is 10 µm. No visible difference could be seen before and after annealing. (b) Raman spectra of a pristine monolayer (red), pristine multilayer (green), and multilayers that were annealed in argon at 550°C (blue). The grey and yellow highlighted boxes denotes $E_{2g}$ and 2LA mode, and $A_{1g}$ vibration mode of $WS_2$. (c-e) XPS spectra of $WS_2$ annealed at 550°C in an Argon atmosphere, showing spectral regions that contain the $O_{1s}$, $S_{2p}$ and $W_{4f}$ peaks, respectively. In (c) the presence of a $WO_x$ phase is clearly observed.

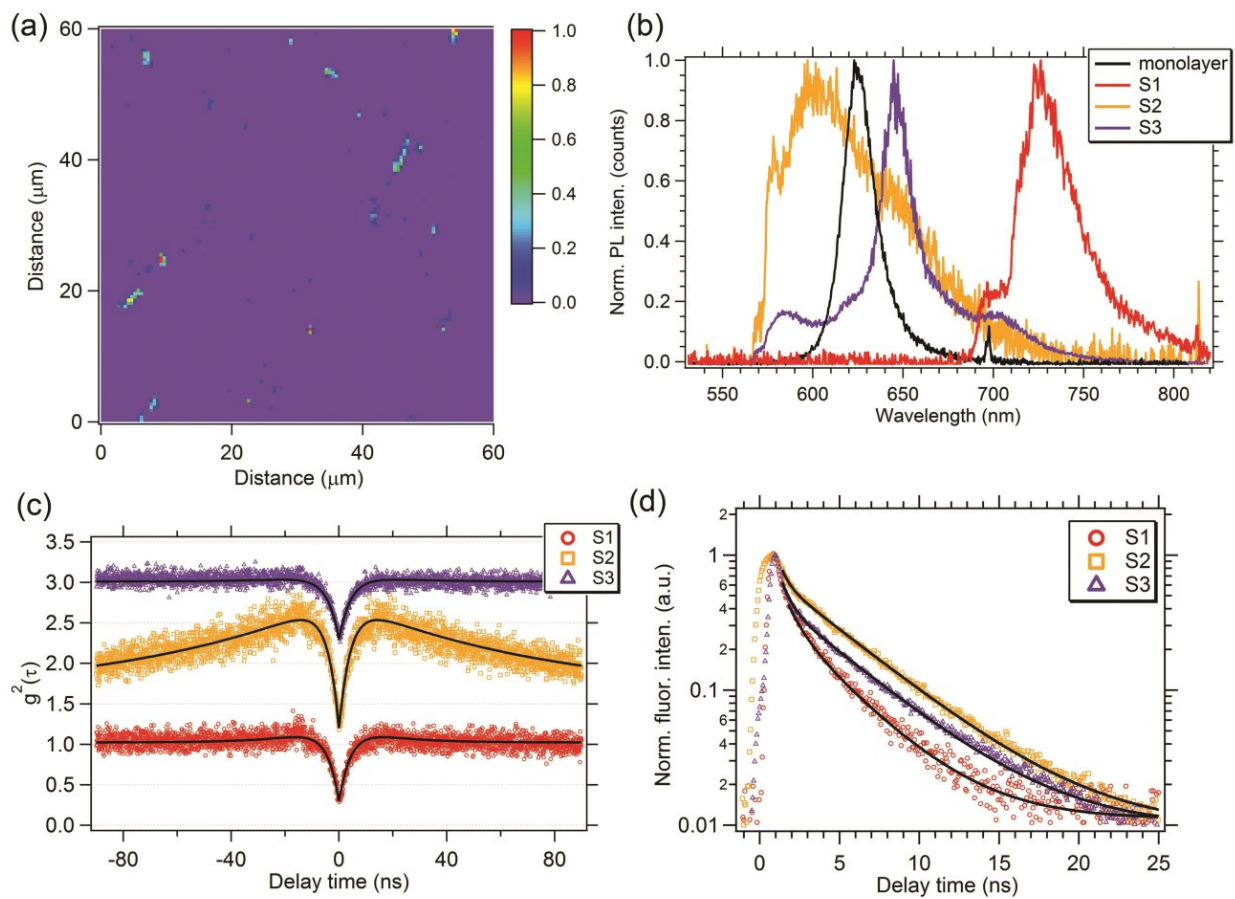

**Figure 2. Single photon emission from annealed multilayers.** (a) A typical confocal photoluminescence map showing several bright spots corresponding to localized defects. (b) Photoluminescence spectra taken from three bright spots. A spectrum from monolayer $WS_2$ is plotted for comparison (c) Second order autocorrelation measurement obtained from the three emitters. The curves are offset vertically for clarity. (d) Time-resolved photoluminescence measurement recorded from the three emitters, yielding excited state lifetimes of 3.5 ns, 4.6 ns, and 4.4 ns, respectively for emitters S1, S2 and S3. The pump power used in (a, b, c) was 300 µW at 532 nm while the pulsed measurement was done using a 512 nm laser (10 MHz, 50 µW).

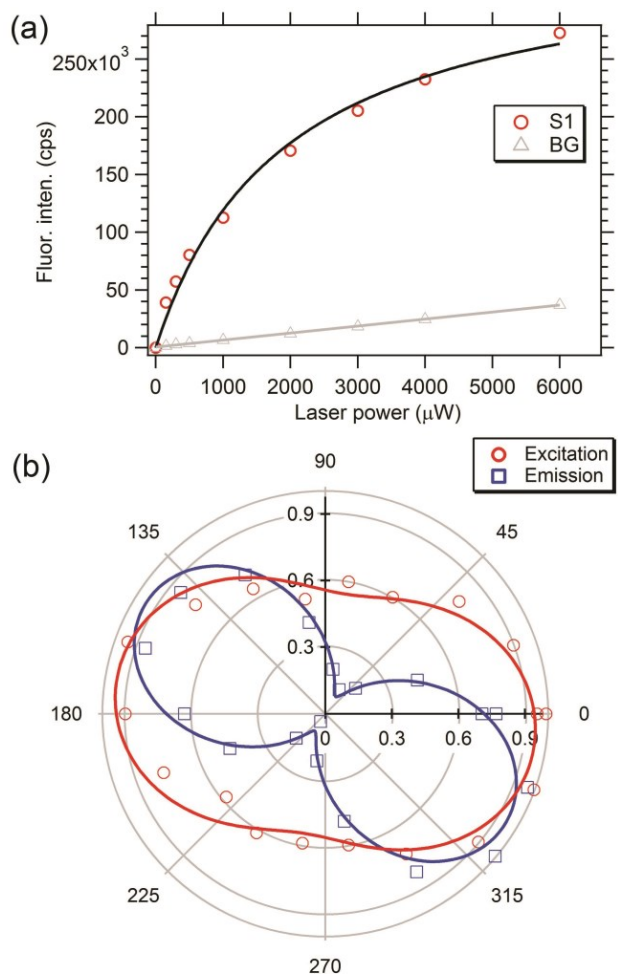

**Figure 3. Fluorescence saturation and polarization of emitter S1.** (a) Fluorescence intensity as a function of pump power. The red open circles and grey open triangles denote background-corrected fluorescence profile of emitter S1 and background fluorescence taken at an area adjacent to the emitter (in Figure 2). The solid lines are corresponding fitted curves. The saturated intensity is 347,000 cps with the saturated pump at 1.9 mW. (b) Excitation (red open circles) and emission (blue open squares) measurement for emitter S1. Solid lines are corresponding fits.

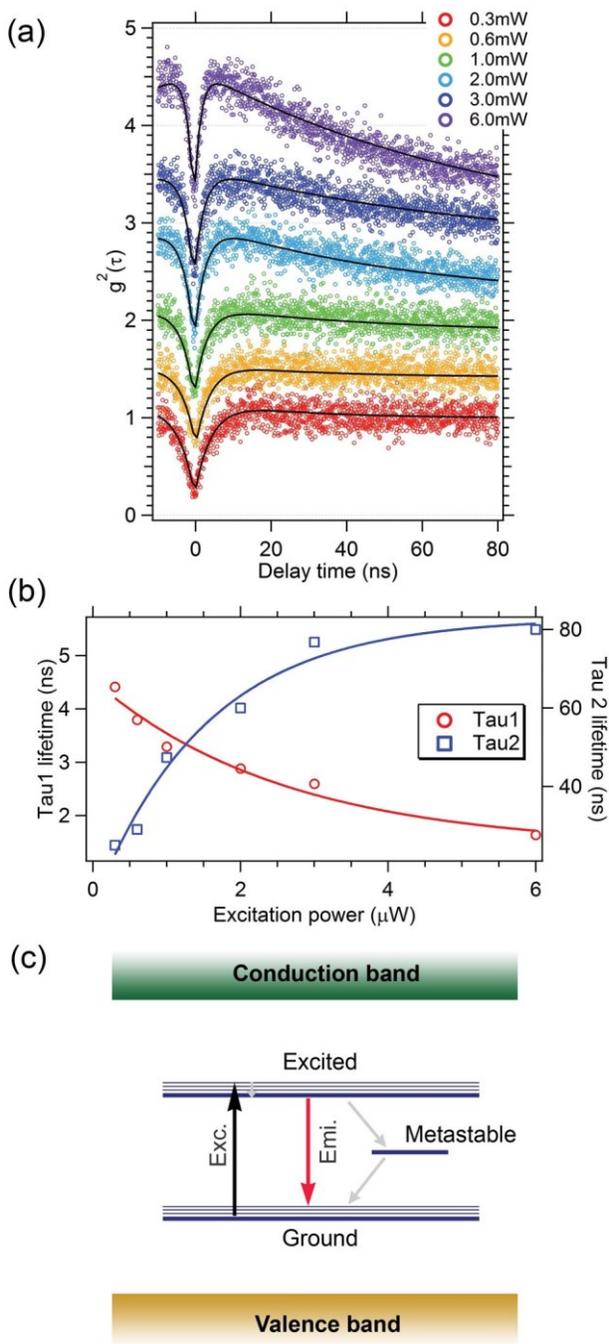

**Figure 4.** (a) Antibunching curves as a function of pump power. Solid lines are fitted profile using the standard three-level model. (b) Emission ($\tau_1$) and metastable ($\tau_2$) lifetime plotted as a function of excitation power. By applying linear fitting and extrapolating the fits to vanishing excitation power, emission ($\tau_1$) and metastable ($\tau_2$) lifetime are calculated to be 4.5 ns and 9.3 ns, respectively. (c) Proposed three-level diagram of the emitters with a ground state, an excited state

and a metastable state. Black, red and grey arrows represent excitation, emission and non-radiative transitions.